\documentclass[12pt]{article}
\usepackage{amscd,amssymb,amsmath,latexsym,enumerate}
\usepackage[mathscr]{euscript}
\usepackage{epsfig}
\usepackage{fancybox}
\usepackage{verbatim}
\usepackage{stmaryrd}

\usepackage[small,nohug]{diagrams}
\diagramstyle[labelstyle=\scriptstyle]

\usepackage{color}

\title{Spectral flow argument localizing an odd index pairing}

\author{Terry A. Loring$^1$ and Hermann Schulz-Baldes$^2$
\\
\\
\\
{\small $^1$Department of Mathematics and Statistics, University of New Mexico, USA}
\\
{\small $^2$Department Mathematik, Friedrich-Alexander-Universit\"at Erlangen-N\"urnberg, Germany}
}

\date{ }

\newcommand{\CM}{{\mathbb C}}

\newcommand{\RM}{{\mathbb R}}

\newcommand{\ZM}{{\mathbb Z}}

\newcommand{\Hh}{{\cal H}}

\newcommand{\one}{{\bf 1}}

\newcommand{\SF}{{\rm Sf}} 
 
\newcommand{\Ind}{{\rm Ind}}

\newcommand{\Ran}{{\rm Ran}} 
 
\newcommand{\Sig}{{\rm Sig}} 
\newcommand{\diag}{{\rm diag}}

\newcommand{\SLoc}{L}

\newcommand{\TapeG}{G_\rho}

\begin{document}

\maketitle

\begin{abstract}
An odd Fredholm module for a given invertible operator on a Hilbert space is specified by an unbounded so-called Dirac operator with compact resolvent and bounded commutator with the given invertible. Associated to this is an index pairing in terms of  a Fredholm operator with Noether index. Here it is shown by a spectral flow argument how this index can be calculated as the signature of a finite dimensional matrix called the spectral localizer. \hfill MSC2010: 19K56, 46L80
\end{abstract}

\vspace{-.8cm}

\section{Statement of the result}
\label{sec-IndPairings}

In a recent paper \cite{LS}, the authors proved that the odd index pairing of an invertible operator on the Hilbert space over an odd dimensional torus with a suitable odd Fredholm module can be calculated as the signature of a finite dimensional matrix called the spectral localizer. This makes it possible to access the index numerically in many cases. The proof in \cite{LS} combined analytic estimates showing that this signature is well-defined with $K$-theoretic arguments on fuzzy spheres. For the special case of a one-dimensional system, it was also shown how the $K$-theoretic part of the proof can be replaced by a spectral flow argument using the $\eta$-invariant and its connection to spectral flow \cite{CP}. It is the aim of this short note to provide a more direct, simple and to our taste simply beautiful argument connecting an odd index pairing to the spectral localizer. It avoids both $K$-theory as  well as the $\eta$-invariant, and instead uses only basic analytical properties of the spectral flow as given in  \cite{Phi,Ph1,DS2}, and recalled in appendix. It is hence of purely functional analytic nature.

\vspace{.1cm}

Let us begin by stating the main result. It extends the main statement of \cite{LS} to the natural and general framework of non-commutative index theory \cite{Con,GVF}. Let $A$ be an invertible operator on a separable Hilbert space $\Hh$. An unbounded odd Fredholm module for $A$ is a selfadjoint invertible (Dirac) operator $D$ with compact resolvent such that the commutator $[A,D]$ extends to a bounded operator (more traditionally, one requires commutators for a dense subset in a C$^*$-algebra containing $A$ to have this property). Associated to $D$ is a so-called Hardy projection $\Pi=\chi(D>0)$ where $\chi$ denotes the indicator function.  Then it is well-known, {\it e.g.}\ \cite{Con} or p.\ 462 in \cite{GVF}, that $[\Pi,A]$ is compact and the Toeplitz operator
\begin{equation}
\label{eq-OddPair}
T\;=\;\Pi \,A\,\Pi+(\one-\Pi)
\;,
\end{equation}
is a bounded Fredholm operator on $\Hh$. Its associated Noether index is denoted by $\Ind(T)$. The operator $T$ and its index are called the index pairing of (the $K_1$-class of) $A$ with (the odd Fredholm module specified by) $D$. In this situation,  the spectral localizer is defined as the operator
$$
\SLoc_{\kappa}
\;=\;
\begin{pmatrix} \kappa\,D & A \\ A^* & -\kappa\,D  \end{pmatrix}
\;,
$$
acting on $\Hh\oplus\Hh$ where $\kappa>0$ is a tuning parameter to be specified later on. It measures the size w.r.t. $A$, which is supposed to satisfy $\|A\|\geq 1$. We will consider finite volume restrictions of the spectral localizer. For given $\rho>0$, let us set
\begin{equation}
\label{eq-Hfinite}
\Hh_\rho
\;=\;
\Ran\big(\chi(D^2\leq \rho^2)\big)
\;.
\end{equation}
As $D$ has compact resolvent, each $\Hh_\rho$ is finite-dimensional.  The surjective partial isometry from $\Hh$ to $\Hh_\rho$ will be denoted by $\pi_\rho$. For any operator $T$ on $\Hh$, we then set $T_\rho=\pi_\rho T\pi^*_\rho$ which is an operator on $\Hh_\rho$. This corresponds to restricting $T$ with Dirichlet boundary conditions. Let us also use $\one_\rho=\pi_\rho\pi_\rho^*$ for the identity on $\Hh_\rho$. We also write $\pi_\rho$ and $\one_\rho$ for the surjective partial isometry and identity on $\Hh_\rho\oplus\Hh_\rho$. With these notations, the finite volume spectral localizer on $\Hh_\rho\oplus\Hh_\rho$ is
\begin{equation}
\label{eq-SLocFinite}
\SLoc_{\kappa,\rho}
\;=\;
\pi_\rho\,L_\kappa\,\pi_\rho^*
\;=\;
\begin{pmatrix} \kappa\,D_\rho & A_\rho \\ A_\rho^* & -\kappa\,D_\rho  \end{pmatrix}
\;.
\end{equation}
The following was proved in \cite{LS} for the special case of $\Hh=\ell^2(\ZM^d,\CM^N)$ with $d$ odd and $D$ built from the components of the position operator on $\ell^2(\ZM^d,\CM^N)$, see the example below.

\vspace{.2cm}

\noindent {\bf Theorem}
{\sl Let $g=\|A^{-1}\|^{-1}>0$ be the gap of an invertible bounded operator $A$ on a separable Hilbert space. Let $D$ specify an odd Fredholm module for $A$. Set
\begin{equation}
\label{eq-MainCond1}
\kappa_0\;=\;\frac{g^3}{12\,\|A\|\,\|[D,A]\|}
\;.
\end{equation}
Suppose that $\kappa$ and $\rho$ are such that 
\begin{equation}
\label{eq-MainCond2}
\frac{2\,g}{\kappa_0}\;\leq\;\frac{2\,g}{\kappa}\;<\;\rho
\;.
\end{equation}
Then the matrix $\SLoc_{\kappa,\rho}$  satisfies the bound
\begin{equation}
\label{eq-GapBound}
(\SLoc_{\kappa,\rho})^2
\;\geq\;
\frac{g^2}{4}\,\one_\rho
\;.
\end{equation}
In particular, $\SLoc_{\kappa,\rho}$ is invertible  and thus has a well-defined signature $\Sig(\SLoc_{\kappa,\rho})$. Then
\begin{equation}
\label{eq-IndSigLink}
\Ind\big(\Pi \,A\,\Pi+(\one-\Pi)\big)
\;=\;
\frac{1}{2}\,
\Sig(\SLoc_{\kappa,\rho})
\;.
\end{equation}
}

\vspace{.3cm}

The main hypothesis is that the commutator $[D,A]$ is bounded, that is, $A$ is differentiable in the non-commutative sense. Then one can choose sufficiently small $\kappa$  and large $\rho$ such that both inequalities in \eqref{eq-MainCond2} hold. The identity \eqref{eq-IndSigLink} then allows to determine the index pairing numerically in many interesting situations, see \cite{Lor,LS} which is based on the example below.  

\vspace{.2cm}

Let us note that $\kappa=0$ forces $\rho=\infty$. Indeed, $L_{0,\rho}$ has vanishing signature for any finite $\rho$. Thus $L_{\kappa,\rho}$ is not invertible for some $\kappa<2g/\rho$. Actually the spectral flow for $\kappa\in[0,2g/\rho]\mapsto L_{\kappa,\rho}$ is equal to the half-signature on the r.h.s.\ of \eqref{eq-IndSigLink}. However, our proof of the Theorem is not based on this spectral flow, but rather starts out with Phillips' Theorem \cite{Ph1} expressing the index pairing on the l.h.s.\ of \eqref{eq-IndSigLink} in terms of a suitable spectral flow, see Section~\ref{sec-proof} below.

\vspace{.2cm}

Another comment on the Theorem concerns the choice of the finite dimensional Hilbert space $\Hh_\rho$. The proof below shows that any other finite subspace of $\Hh$ may be used without altering the signature, provided that it contains $\Hh_\rho$ for some $\rho$ satisfying \eqref{eq-MainCond2}. For the example below this means that one can choose finite volumes to be balls or cubes or triangles, etc.

\vspace{.3cm}

\noindent {\bf Example:}  Let $d$ be odd. The Hilbert space is $\ell^2(\ZM^d)$ over a $d$-dimensional lattice $\ZM^d$, extended by a fiber $\CM^N$. Thus $\Hh=\ell^2(\ZM^d,\CM^N)$. On this Hilbert space act the $d$ components $X_1,\ldots,X_d$ of the selfadjoint commuting position operators defined by $X_j|n\rangle=n_j|n\rangle$ where $n=(n_1,\ldots,n_d)\in\ZM^d$ and $|n\rangle\in\ell^2(\ZM^d)$ is the Dirac Bra-Ket notation for the unit vector localized at $n$. Suppose furthermore that $\gamma_1,\ldots,\gamma_{d}$ is a self-adjoint irreducible representation of the Clifford algebra $\CM_{d}$ on $\CM^N$. This determines $N=2^{\frac{d-1}{2}}$ in terms of $d$. From this data, the selfadjoint Dirac operator  is
%
$$
D\;=\;\sum_{j=1}^{d}X_j\otimes\gamma_j
\;+\;
|0\rangle\langle 0|\otimes \gamma_1
\;.
$$
%
The last summand is added to ensure that $D$ is invertible. Clearly, $D$ has a compact resolvent. It defines a Fredholm module for $A$ if these operators have a bounded commutator 
$$
\|[D,A]\|\,<\,\infty
\;.
$$
Now for $d=1$, $\Pi=\chi(D>0)$ is the classical Hardy projection and, if furthermore $A$ is periodic, $T$ is the classical Toeplitz operator (strictly speaking extended by $\one-\Pi$ to the full Hilbert space). Its index is then known to be the winding number of $A$ which according to \eqref{eq-IndSigLink} can hence be calculated readily from the spectral localizer. For higher odd dimension $d$, the index pairing plays an important role in the theory of topological insulators, {\it e.g.}\  \cite{Lor,GS,PSB}.
\hfill $\diamond$

\vspace{.5cm}

\noindent{\bf Acknowledgments:} The authors thank the Simons Foundation (CGM	419432), the NSF (DMS 1700102)  and the DFG (SCHU 1358/3-4) for financial support.

\section{The proof of the constancy of the signature}
\label{sec-proof0}

The first step of the proof, namely that \eqref{eq-GapBound} follows from \eqref{eq-MainCond2}, is preparatory. It will also be shown in this section that the signature is the same for all pairs $\kappa,\rho$ satisfying \eqref{eq-MainCond2}. While this is already contained in \cite{LS}, the argument given here is simpler and improves the estimates slightly. The proof will use an even and differentiable tapering function $G_\rho:\RM\to [0,1]$ with three properties: 
\begin{itemize}
\item[{\rm (i)}] $G_\rho(x)=1$ for $|x|\leq\frac{\rho}{2}$; 
\item[{\rm (ii)}]  $G_\rho(x)=0$ for $|x|\geq\rho$; 
\item[{\rm (iii)}]  The Fourier transform $\widehat{G'_\rho}$ of the derivative $G'_\rho$ has an $L^1$-norm bounded by $8 \rho^{-1}$. 
\end{itemize}
Such a function is explicitly constructed in \cite{LS} where it is also shown (using \cite[Lemma~10.15]{GVF}) that 
$$
\|[G_\rho(D),A]\|\;\leq\;8\,\rho^{-1}\,\|D,A]\|
\;.
$$
Setting $D'=\begin{pmatrix} D & 0 \\ 0 & -D \end{pmatrix}$ and $H=\begin{pmatrix} 0 & A \\ A^* & 0 \end{pmatrix}$, one has $D'H+HD'=\begin{pmatrix} 0 & [D,A] \\ [D,A]^* & 0 \end{pmatrix}$ and, due to $G_\rho(-D)=G_\rho(D)$, also
\begin{equation}
\|[G_\rho(D'),H]\|\;\leq\;8\,\rho^{-1}\,\|[D,A]\|
\;.
\label{eq-TapeCommu}
\end{equation}
In order to connect radii $\rho$ and $\rho'\geq \rho$, let us introduce
$$
L_{\kappa,\rho,\rho'}(\lambda)
\;=\;
\kappa\;\pi_{\rho'}\,D'\,\pi_{\rho'}^{*}
\;+\;
\pi_{\rho'}G_{\lambda,\rho}\,H\, G_{\lambda,\rho}\,\pi_{\rho'}^{*}
\;,
$$
acting on $\Hh_{\rho'}\oplus\Hh_{\rho'}$ where $0\leq\lambda\leq1$ and
$$
G_{\lambda,\rho} \;=\; (1-\lambda)\one\;+\;\lambda\, G_{\rho}(D')
\;.
$$
Also \eqref{eq-MainCond2} is supposed to hold. Notice that $L_{\kappa,\rho,\rho'}(0) = L_{\kappa,\rho'}$. The first goal is to show that $L_{\kappa,\rho,\rho'}(\lambda)$ is always invertible and that
its square is bounded below by $\frac{g^{2}}{4}\one_{\rho'}$ when $\lambda=0$. The square of $L_{\kappa,\rho,\rho'}(\lambda)$ becomes
$$
L_{\kappa,\rho,\rho'}(\lambda)^{2}
\;=\;
\kappa^{2}\,
\pi_{\rho'}(D')^{2}\pi_{\rho'}^{*}
\,+\,
\left(\pi_{\rho'}G_{\lambda,\rho}\,H\, G_{\lambda,\rho}\pi_{\rho'}^{*}\right)^{2}
\,+\,
\kappa\,\pi_{\rho'}G_{\lambda,\rho}(D'H+HD') G_{\lambda,\rho}\pi_{\rho'}^{*}
\;.
$$
Let us begin with the second summand:
\begin{align*}
\big(  \pi_{\rho'} G_{\lambda,\rho}\,H\, G_{\lambda,\rho}\pi_{\rho'}^{*}\big)^{2}
 & =\,\pi_{\rho'}G_{\lambda,\rho}\,H\,G_{\lambda,\rho}^{2}\,H\,G_{\lambda,\rho}\pi_{\rho'}^{*}\\
 & \geq\,\pi_{\rho'}G_{\lambda,\rho}\,H\,  G_{\rho}(D')^{2}\,H\,G_{\lambda,\rho}\pi_{\rho'}^{*}\\
 & =\,\pi_{\rho'}G_{\lambda,\rho}G_{\rho}(D')\,H^2\,G_{\rho}(D')G_{\lambda,\rho}\pi_{\rho'}^{*}+\pi_{\rho'}G_{\lambda,\rho}\left[G_{\rho}(D')\,H,\left[G_{\rho}(D'),H\right]\right]G_{\lambda,\rho}\pi_{\rho'}^{*}\\
 & >\,g^{2}\pi_{\rho'}G_{\lambda,\rho}^{2}G_{\rho}(D')^{2}\pi_{\rho'}^{*}+\pi_{\rho'}G_{\lambda,\rho}\left[G_{\rho}(D')\, H,\left[G_{\rho}(D'),H\right]\right]G_{\lambda,\rho}\pi_{\rho'}^{*}\\
 & \geq \, g^{2}\pi_{\rho'}G_{\rho}(D')^{4}\pi_{\rho'}^{*}+\pi_{\rho'}G_{\lambda,\rho}\left[G_{\rho}(D')H,\left[G_{\rho}(D'),H\right]\right]G_{\lambda,\rho}\pi_{\rho'}^{*}
 \;.
\end{align*}
For the special case of $\lambda=0$ one has the better estimate
$$
\left(\pi_{\rho'}G_{0,\rho}\,H\, G_{0,\rho}\pi_{\rho'}^{*}\right)^{2} 
\;\geq\;
g^{2}\pi_{\rho'}G_{\rho}(D')^{2}\pi_{\rho'}^{*}+\pi_{\rho'}G_{\lambda,\rho}\left[G_{\rho}(D')\,H\,,\left[G_{\rho}(D'),H\right]\right]G_{\lambda,\rho}\pi_{\rho'}^{*}
\;.
$$
Furthermore, by spectral calculus of $D'$ one has the bound 
$$
\kappa^2\, \pi_{\rho'}(D')^2\pi_{\rho'}^*
\;\geq\;
g^2\,\pi_{\rho'}(\one-\TapeG(D')^2)\pi_{\rho'}^*
\;,
$$
because the bound  holds for spectral parameters in $[\frac{1}{2}\rho,\rho']$ due to \eqref{eq-MainCond2} and $\one-\TapeG(D')^2\leq \one$, while it holds trivially on $[0,\frac{1}{2}\rho]$. Since
$$
\one-G_{\rho}(D')^{2}+G_{\rho}(D')^{4}
\;\geq\;\tfrac{3}{4}\,\one
\;,
$$
it thus follows
$$
L_{\kappa,\rho,\rho'}(\lambda)^{2}
\,\geq\,
\tfrac{3}{4}\,g^{2}\one_{\rho'}
+
\pi_{\rho'}G_{\lambda,\rho}\big(\left[G_{\rho}(D')\,H,\left[G_{\rho}(D'),H\right]\right]+\kappa(D'H+HD')\big)G_{\lambda,\rho}\pi_{\rho'}^{*}
\,,
$$
and in the special case $\lambda=0$,
$$
L_{\kappa,\rho,\rho'}(0)^{2}
\,\geq\,
g^{2}\,\one_{\rho'}
+
\pi_{\rho'}G_{\lambda,\rho}\big(\left[G_{\rho}(D')\,H,\left[G_{\rho}(D'),H\right]\right]+\kappa(D'H+HD')\big)G_{\lambda,\rho}\pi_{\rho'}^{*}
\,.
$$
Finally the error term is bounded using the tapering estimate \eqref{eq-TapeCommu}:
\begin{align*}
\big\|
[\TapeG(D')\, H ,[\TapeG(D') ,H]]
\,+ \,\kappa(HD'+D'H)
\big\|
& \;\leq\;
\Big(
2\,\|\TapeG(D')\, H \|\,8(\rho)^{-1}\,+\,\kappa
\Big)
\|[A ,D]\|
\\
&
\;<\;
\Big(
\|A\|\,8\,(g)^{-1}\,+\,1\Big)\,\kappa\,\|[A ,D]\|
\\
&
\;\leq\;
\|A\|\,9 \,g^{-1}\,\kappa\,\|[A ,D]\|
\\
&
\;\leq\;\tfrac{3}{4}\,g^2
\;,
\end{align*}
where the second inequality used the second inequality in  \eqref{eq-MainCond2} as well as $\|\TapeG(D') \|\leq 1$, the third one $\|A\|\geq 1$ and $g\leq 1$, and finally the last inequality follows from the first inequality in \eqref{eq-MainCond2}. Together one infers $L_{\kappa,\rho,\rho'}(\lambda)^{2}>0$ and $L_{\kappa,\rho,\rho'}(0)^{2}\geq\frac{1}{4}\,g^2\,\one_{\rho'}$. 

\vspace{.1cm}

Finally, let us show that
$$
\Sig\left(L_{\kappa,\rho}\right)
\;=\;
\Sig\left(L_{\kappa',\rho'}\right)
\;,
$$
for pairs $\kappa,\rho$ and $\kappa',\rho'$ in the permitted range of parameters.  Without loss of generality let $\rho\leq\rho'$. Clearly $L_{\kappa,\rho}$ is continuous in $\kappa$,  a homotopy argument allows to
reduce to the case $\kappa=\kappa'$. Thus one needs to show 
$$
\Sig\left(L_{\kappa,\rho,\rho}(0)\right)
\;=\;
\Sig\left(L_{\kappa,\rho,\rho'}(0)\right)
\;,
$$
when $\rho\leq\rho'$ and \eqref{eq-MainCond2} is true for $\kappa$ and $\rho$. Clearly $L_{\kappa,\rho,\rho'}(\lambda)$ is continuous in $\lambda$, so it suffices to prove 
$$
\Sig\left(L_{\kappa,\rho,\rho}(1)\right)
\;=\;
\Sig\left(L_{\kappa,\rho,\rho'}(1)\right)
\;.
$$
Consider 
$$
L_{\kappa,\rho,\rho'}(1)
\;=\;
\kappa\pi_{\rho'}D'\pi_{\rho'}^{*}
+\pi_{\rho'}G_{\rho}(D')\,H\, G_{\rho}(D')\pi_{\rho'}^{*}
\;.
$$
Now $D'$ commutes with $\pi_{\rho'}^*\pi_{\rho'}$ so that $L_{\kappa,\rho,\rho'}(1)$ decomposes into a direct sum. Let $\pi_{\rho',\rho}=\pi_{\rho'}\ominus\pi_{\rho}$ be the surjective partial isometry onto $(\Hh\oplus\Hh)_{\rho'}\ominus (\Hh\oplus\Hh)_{\rho}$. Then
$$
L_{\kappa,\rho,\rho'}(1)
\;=\;
L_{\kappa,\rho,\rho}(1)\oplus\,\pi_{\rho',\rho}\,\kappa\,D'\,\pi_{\rho',\rho}^{*}
\;.
$$
The signature of $\pi_{\rho',\rho}\,D'\,\pi_{\rho',\rho}^{*}$ vanishes so that
$$
\Sig(L_{\kappa,\rho,\rho'}(1))\;=\;\Sig(L_{\kappa,\rho,\rho}(1))
\;.
$$
%

\section{The proof of index$\, =\, $half-signature via spectral flow}
\label{sec-proof}

Now let us turn to the main objective of this note, namely the proof of \eqref{eq-IndSigLink} using only the well-known properties of the spectral flow listed in the appendix. Due to the stability result of the signature proved in the last section, it is therefore sufficient to show that for some $\kappa$ and $\rho$ satisfying \eqref{eq-MainCond2} the equality \eqref{eq-IndSigLink}  holds. In the argument below, we will first choose $\kappa$ sufficiently small, and then $\rho$ sufficiently large. The starting point is the fundamental connection, described in the Appendix, between an index pairing and the spectral flow: 
\begin{equation}
\label{eq-IndOddSF}
\Ind(T)
\;=\;
\Ind(\Pi U\Pi+\one-\Pi)
\;=\;
\SF(U^*(2\Pi-\one)U,2\Pi-\one)
\;.
\end{equation}
Here the unitary $U=A|A|^{-1}$ is the polar of $A$. Deforming $2\Pi-\one$ into $\kappa\,D$ does not lead to spectral flow and therefore
\begin{equation}
\label{eq-IndOddSFx}
\Ind(T)
\;=\;
\SF(\kappa\, U^*DU,\kappa\,D)
\;.
\end{equation}
On the r.h.s.\ appears the spectral flow along the straight-line path $t\in[0,1]\mapsto (1-t)\kappa \, U^*DU+t\,\kappa\,D$ between two unbounded operators. As $U^*DU-D=U^*[D,U]$ is bounded, it is intuitively clear that this only invokes low lying spectrum of $D$. Technically, the spectral flow between unbounded operators can be defined (see the Appendix) by replacing $D$ by an increasing bounded function of $D$ such as $\tanh(D)$. To be precise, let $F_\rho:\RM\to\RM$ be an increasing smooth function with $F_\rho(x)=x$ for $|x|\leq \rho$ and $F_\rho(x)=2\rho=-F_\rho(-x)$ for $x\geq 2\rho$. Then 
$$
\SF(\kappa\, U^*DU,\kappa\,D)
\;=\;
\SF(\kappa\, U^*F_\rho(D)U,\kappa\,F_\rho(D))
\;,
$$
where on the r.h.s.\ appears the spectral flow for the straight line between two bounded operators. Note that indeed $F_\rho(D)-2\rho(2\Pi-\one)$ is compact so that also the difference $U^*F_\rho(D)U-F_\rho(D)$ is compact, and hence the straight line from $\kappa\, U^*F_\rho(D)U$ to $\kappa\,F_\rho(D)$ is indeed inside the selfadjoint Fredholm operators. Furthermore, this straight-line path can be deformed into the one on the r.h.s.\ of \eqref{eq-IndOddSF} (still within the selfadjoint Fredholm operators) which shows  \eqref{eq-IndOddSFx}. In the following, one should strictly speaking always replace $D$ by $F_\rho(D)$. As only the equality $F_\rho(D)=D$ on $\Hh_\rho$ is relevant in the following, we decided to stick with the formulation with unbounded operators for sake of clarity of the argument.

\vspace{.1cm}

Hence let us continue from \eqref{eq-IndOddSFx}. Using the additivity of the spectral flow leads to
\begin{align}
\Ind(T)
& \;=\;
\SF
\left(
\begin{pmatrix} U & 0 \\ 0 & \one\end{pmatrix}^*
\begin{pmatrix} \kappa\,D & 0\\ 0 & -\kappa\,D \end{pmatrix}
\begin{pmatrix} U & 0 \\ 0 & \one\end{pmatrix}
,
\begin{pmatrix} \kappa\,D & 0 \\ 0 & -\kappa\,D \end{pmatrix}
\right)
\nonumber
\\
& 
{
\;=\;
\SF
\left(
\begin{pmatrix} \kappa\,D & \one \\ \one & -\kappa\,D \end{pmatrix}
,
\begin{pmatrix} U & 0 \\ 0 & \one\end{pmatrix}
\begin{pmatrix} \kappa\,D & 0\\ 0 & -\kappa\,D \end{pmatrix}
\begin{pmatrix} U & 0 \\ 0 & \one\end{pmatrix}^*
\right)
}
\nonumber
\\
& 
\;=\;
\SF
\left(
{
\begin{pmatrix} \kappa\,D & 0\\ 0 & -\kappa\,D \end{pmatrix}
,
\begin{pmatrix} U & 0 \\ 0 & \one\end{pmatrix}
\begin{pmatrix} \kappa\,D & \one \\ \one & -\kappa\,D \end{pmatrix}
\begin{pmatrix} U & 0 \\ 0 & \one\end{pmatrix}^*
}
\right)
\nonumber
\\
& \;=\;
\SF
\left(
{
\begin{pmatrix} \kappa\,D & 0 \\ 0 & -\kappa\,D \end{pmatrix}
,
\begin{pmatrix} \kappa\,UDU^* & U \\ U^* & -\kappa\,D \end{pmatrix}
}
\right)
\;,
\label{eq-intermed}
\end{align}
where in the second to last step a mass term was added. This only further opens the spectral gap of the end points of the path and hence does not modify the spectral flow. 

\vspace{.1cm}

Now the {right} entry in \eqref{eq-intermed} of the spectral flow in the last equation is essentially the spectral localizer of $U$ w.r.t.\ $D$, provided that one can replace the upper left entry by $\kappa D$. The spectral localizer of $U$ has a gap by the argument above (which also covers the case $\rho=\infty$). As {$UDU^*=D-[D,U]U^*$} and $[D,U]$ is bounded, this gap does not close along the linear path connecting {$UDU^*$} to $D$ for $\kappa$ sufficiently small, so that also the above left entry remains invertible, namely
$$
\SF
\left(
\begin{pmatrix} \kappa\,{UDU^*} & U \\ U^* & -\kappa\,D \end{pmatrix}
,
\begin{pmatrix} \kappa\,D & U \\ U^* & -\kappa\,D \end{pmatrix}
\right)
\;=\;
0
\;,
$$
where again the straight line path between the arguments  is taken. Due to the concatenation property of the spectral flow one concludes that, for $\kappa$ sufficiently small
$$
\Ind(T)
\;=\;
\SF
\left(
{
\begin{pmatrix} \kappa\,D & 0 \\ 0 & -\kappa\,D \end{pmatrix}
,
\begin{pmatrix} \kappa\,D & U \\ U^* & -\kappa\,D \end{pmatrix}
}
\right)
\;.
$$
Deforming $U$ to $A$ via polar decomposition shows
$$
\Ind(T)
\;=\;
\SF
\left(
{
\begin{pmatrix} \kappa\,D & 0 \\ 0 & -\kappa\,D \end{pmatrix}
,
\begin{pmatrix} \kappa\,D & A \\ A^* & -\kappa\,D \end{pmatrix}
}
\right)
\;.
$$
The {right} entry is the spectral localizer. The final step is to localize this formula on $\Hh_\rho$ in the decomposition $\Hh=\Hh_\rho\oplus\Hh_{\rho^c}$. Let $\pi_{\rho^c}$ be the surjective partial isometry onto $\Hh_{\rho^c}$. For any operator $A$, let $A_\rho=\pi_\rho A\pi_\rho^*$ and $A_{\rho^c}=\pi_{\rho^c} A(\pi_{\rho^c})^*$ be the two restrictions with Dirichlet boundary conditions. The Dirac operator $D$ is diagonal w.r.t.\ this decomposition, that is $D=D_\rho\oplus D_{\rho^c}$. The spectral localizer is not diagonal in this grading, but we will homotopically deform it into something diagonal by
\begin{equation}
\label{eq-OffDiagonalHomotopy}
\SLoc_{\kappa}(t)
\;=\;
\SLoc_{\kappa,\rho}
\oplus
\SLoc_{\kappa,\rho^c}
\,+\,
t\,
\begin{pmatrix}
 0 & \pi_{\rho} H(\pi_{\rho^c})^*
\\
 \pi_{\rho^c} H\pi^*_{\rho} & 0 
\end{pmatrix}
\;.
\end{equation}
Note that the perturbation on the r.h.s. of \eqref{eq-OffDiagonalHomotopy} is compact. For $t=1$ one has $\SLoc_{\kappa}(1)=\SLoc_{\kappa}$. We will now show that the path $t\in[0,1]\mapsto \SLoc_{\kappa}(t)$ is within the invertibles, provided $\rho$ is sufficiently large. Indeed, $\SLoc_{\kappa,\rho^c}$ is invertible and has a central gap around $0$ growing linearly in $\rho$ because
\begin{align}
(\SLoc_{\kappa,\rho^c})^2
& 
\;=\;
\begin{pmatrix}
\kappa^2 D_{\rho^c}^2\,+\,A_{\rho^c}(A_{\rho^c})^* & \kappa(D_{\rho^c}A_{\rho^c}-A_{\rho^c}D_{\rho^c})
\\
\kappa(D_{\rho^c}A_{\rho^c}-A_{\rho^c}D_{\rho^c})^* & \kappa^2 D_{\rho^c}^2\,+\,(A_{\rho^c})^*A_{\rho^c}
\end{pmatrix}
\nonumber
\\
&\;\geq\;
{
(\kappa^2\rho^2\,\one_2\;-\;\kappa\|[D,A]\|)\,\one_{\rho^c}}
\;.
\label{eq-InequHel}
\end{align}
This remains strictly positive and actually larger than $\frac{1}{2}\kappa^2\rho^2$ if \eqref{eq-MainCond1} holds. Moreover, the other diagonal entry $\SLoc_{\kappa,\rho}$ has a gap bounded below by {$\frac{g}{2}$}. Hence $\SLoc_{\kappa}(t)$ is given by
$$
|\SLoc_{\kappa,\rho}
\oplus
\SLoc_{\kappa,\rho^c}|^{{\frac{1}{2}}}
\left(
S
\,+\,
t\,
\begin{pmatrix}
 0 & |\SLoc_{\kappa,\rho}|^{-\frac{1}{2}}
{\pi_{\rho}} H{\pi^*_{\rho^c}}|\SLoc_{\kappa,\rho^c}|^{-\frac{1}{2}}
\\
|\SLoc_{\kappa,\rho^c}|^{-\frac{1}{2}}
{\pi_{\rho^c}} H{\pi^*_{\rho}} |\SLoc_{\kappa,\rho}|^{-\frac{1}{2}} & 0 
\end{pmatrix}
\right)
|\SLoc_{\kappa,\rho}
\oplus
\SLoc_{\kappa,\rho^c}|^{{\frac{1}{2}}}
,
$$
where $S$ is a diagonal selfadjoint unitary (diagonal in the direct sum $\Hh=\Hh_\rho\oplus\Hh_\rho^c$). As the off-diagonal entries satisfy
$$
\big\|
|\SLoc_{\kappa,\rho^c}|^{-\frac{1}{2}}
{\pi_{\rho^c}} H{\pi^*_{\rho}} |\SLoc_{\kappa,\rho}|^{-\frac{1}{2}} 
\big\| 
\;\leq\;
\frac{C}{{\sqrt{\kappa\rho g}}}
$$
for some constant $C$, they are smaller than $1$ in norm for $\rho$ sufficiently large. Thus one concludes that that $\SLoc_{\kappa}(t)$ is invertible for all $t\in[0,1]$, provided $\rho$ is sufficiently large.  Consequently, as always for the straight line path,
$$
\SF(
{
\SLoc_{\kappa},\SLoc_{\kappa,\rho}
\oplus
\SLoc_{\kappa,\rho^c}
})
\;=\;
0
\;.
$$
Therefore by concatenation and the additivity of the spectral flow
$$
\Ind(T)
\;=\;
\SF
\left(
{
\begin{pmatrix} \kappa\,D & 0 \\ 0 & -\kappa\,D \end{pmatrix}
,
\SLoc_{\kappa,\rho}
\oplus
\SLoc_{\kappa,\rho^c}
}
\right)
\;=\;
\SF
\left(
{
\begin{pmatrix} \kappa\,D_\rho & 0 \\ 0 & -\kappa\,D_\rho \end{pmatrix}
,\SLoc_{\kappa,\rho}
}
\right)
\;,
$$
because the above inequality \eqref{eq-InequHel} shows that there is no spectral flow on $\Hh_{\rho^c}$. Now for finite dimensional selfadjoint matrices, the spectral flow is the difference of signatures of the matrices divided by $2$. As the signature of $\diag(D_\rho,-D_\rho)$ vanishes, the equality  \eqref{eq-IndSigLink} follows.

\appendix

\section{Review of spectral flow}
\label{app-SFReview}

For the convenience of the reader, we briefly collect the main relevant information from \cite{Phi,Ph1,DS2} on the spectral flow. Let $t\in[0,1]\mapsto F_t$ be a continuous path of bounded selfadjoint Fredholm operators. Then there is an $\epsilon>0$ such that $F_t$ has no essential spectrum in $(-\epsilon,\epsilon)$  for all $t\in[0,1]$. There may, however, be discrete spectrum  (isolated  eigenvalues of finite multiplicity) of $T_t$ in the interval $(-\epsilon,\epsilon)$. Intuitively, the spectral flow is then the number of eigenvalues moving past $0$ in the positive direction minus the number of those eigenvalues moving past $0$ in the negative direction. For an analytic path $t\in[0,1]\mapsto F_t$, a rigorous definition of the spectral flow $\SF(t\in[0,1]\mapsto F_t)$ can immediately spelled out due to analytic perturbation theory \cite{APS}. For merely continuous paths, Phillips give a careful definition of the spectral flow by using spectral projections at discrete times \cite{Phi}. While the reader is referred to \cite{Phi} for the definition, let us recall the main basic properties of the spectral flow that are used in the proof above:
\begin{itemize}

\item[{\rm (i)}] (Homotopy invariance) Let $s\in[0,1]\mapsto F_t(s)$ be a homotopy of paths with fixed end points $F_0(s)$ and $F_1(s)$. Then
$$
\SF(t\in[0,1]\mapsto F_t(0))
\;=\;
\SF(t\in[0,1]\mapsto F_t(1))
\;.
$$

\item[{\rm (ii)}] (Concatenation)  For paths $t\in[0,1]\mapsto F_t$ and $t\in[1,2]\mapsto F_t$,
$$
\SF(t\in[0,1]\mapsto F_t)
\;+\;
\SF(t\in[1,2]\mapsto F_t)
\;=\;
\SF(t\in[0,2]\mapsto F_t)
\;.
$$
\item[{\rm (iii)}] (Unitary invariance) For any unitary $U$,
$$
\SF(t\in[0,1]\mapsto U^*F_tU)
\;=\;
\SF(t\in[0,1]\mapsto F_t)
\;.
$$

\item[{\rm (iv)}] (Additivity)  For paths $t\in[0,1]\mapsto F_t$ and $t\in[0,1]\mapsto F'_t$,
$$
\SF(t\in[0,1]\mapsto F_t\oplus F'_t)
\;=\;
\SF(t\in[0,1]\mapsto F_t)
\;+\;
\SF(t\in[0,1]\mapsto F'_t)
\;.
$$

\item[{\rm (v)}] For a path $t\in[0,1]\mapsto F_t$ with $0$ not in the spectrum $\sigma(F_t)$ of $F_t$ for all $t\in[0,1]$,
$$
\SF(t\in[0,1]\mapsto F_t)
\;=\;
0
\;.
$$
\end{itemize}

The starting point for the proof in Section~\ref{sec-proof} is the connection between spectral flow and index pairings. Let $P$ be a projection and $U$ a unitary such that $[U,P]$ is a compact operator, then $PUP+\one-P$ is a Fredholm operator (with pseudo-inverse $PU^*P+\one-P$). Associated is a selfadjoint Fredholm operator $F=2P-\one$ as well as a path  $t\in[0,1]\mapsto t\,F+(1-t) U^*FU$ of selfadjoint Fredholm operators, with end points $U^*FU$ and $F$. Then
\begin{equation}
\label{eq-SFInd}
\Ind(PUP+\one-P)
\;=\;
\SF(t\in[0,1]\mapsto t\,F+(1-t) U^*FU)
\;.
\end{equation}
The first proof of the equality \eqref{eq-SFInd} seems to be in \cite{Ph1}, an alternative homotopy argument can be found in \cite{DS2}. For sake of brevity, the spectral flow on the r.h.s. \eqref{eq-SFInd} is also denoted by $\SF(U^*FU,F)$.  More generally, given two selfadjoint Fredholm operator $F_0$ to $F_1$ with compact difference $F_1-F_0$, we will always write $\SF(F_0,F_1)$ for the spectral flow of the straight-line path, that is,
$$
\SF(F_0,F_1)
\;=\;
\SF(t\in[0,1]\mapsto (1-t)\,F_0\,+\,t\, F_1)
\;.
$$

\vspace{.2cm}

Finally let us add a comment on the spectral flow of paths $t\in[0,1]\mapsto D_t$ of unbounded selfadjoint Fredholm operators (continuity w.r.t.\ the graph topology \cite{BCP}). One then obtains a path $t\in[0,1]\mapsto F_t=\tanh(D_t)$ of bounded selfadjoint Fredholm operators and can use its spectral flow to define the spectral flow of $t\in[0,1]\mapsto D_t$. Instead of $\tanh$ any increasing smooth function $F$ with $F(0)=0$ and $F'(0)>0$ can be used. All of the above properties naturally transpose to the unbounded case.


\end{document}